\documentclass[twocolumn,prl,floatfix,superscriptaddress,nobibnotes,nolongbibliography]{revtex4-2}

\usepackage{amsmath}
\usepackage{amssymb}
\usepackage{graphicx}
\usepackage{hyperref}
\usepackage[normalem]{ulem}


\usepackage{xcolor}

\hypersetup{
    colorlinks=true,
    linkcolor={blue!80!black},
    citecolor={blue!80!black},
    urlcolor={blue!80!black}
}

\usepackage{color}

\begin{document}

\title{From two dimensions to wire networks in a dice-lattice Josephson array}
\author{J. D. Bondar}
\affiliation{Center for Quantum Devices, Niels Bohr Institute,
University of Copenhagen, 2100 Copenhagen, Denmark}
\affiliation{Department of Physics and Astronomy, Purdue University, West Lafayette, Indiana 47907, USA}
\author{L. Banszerus}
\affiliation{Center for Quantum Devices, Niels Bohr Institute, University of Copenhagen, 2100 Copenhagen, Denmark}
\affiliation{Faculty of Physics, University of Vienna, 1090 Vienna, Austria}
\author{W. Marshall}
\affiliation{Center for Quantum Devices, Niels Bohr Institute,
University of Copenhagen, 2100 Copenhagen, Denmark}
\affiliation{Department of Physics, University of Washington, Seattle, Washington 98195, USA}
\author{T. Lindemann}
\affiliation{Department of Physics and Astronomy, Purdue University, West Lafayette, Indiana 47907, USA}
\author{T.~Zhang}
\affiliation{Department of Physics and Astronomy, Purdue University, West Lafayette, Indiana 47907, USA}

\author{M. J. Manfra}
\affiliation{Department of Physics and Astronomy, Purdue University, West Lafayette, Indiana 47907, USA} 
\affiliation{
Purdue Quantum Science and Engineering Institute, Purdue University,  West Lafayette, Indiana 47907, USA}
\affiliation{Elmore Family School of Electrical and Computer Engineering, Purdue University, West Lafayette, Indiana 47907, USA}
\affiliation{School of Materials Engineering, Purdue University, West Lafayette, Indiana 47907, USA}
\author{C. M. Marcus}
\affiliation{Center for Quantum Devices, Niels Bohr Institute,
University of Copenhagen, 2100 Copenhagen, Denmark}
\affiliation{Materials Science and Engineering, and Department of Physics, University of Washington, Seattle, Washington 98195, USA
}
\author{S. Vaitiekėnas}
\affiliation{Center for Quantum Devices, Niels Bohr Institute,
University of Copenhagen, 2100 Copenhagen, Denmark}
\date{\today}

\begin{abstract} 
We investigate Josephson arrays consisting of a dice-lattice network of superconducting weak links surrounding rhombic plaquettes of proximitized semiconductor. Josephson coupling of the weak links and electron density in the plaquettes are independently controlled by separate electrostatic gates. Applied magnetic flux results in an intricate pattern of switching currents associated with frustration, $f$. For depleted plaquettes, the switching current is nearly periodic in $f$, expected for a phase-only description, while occupied plaquettes yield a decreasing envelope of switching currents with increasing $f$. A model of flux dependence based on ballistic small-area junctions and diffusive large-area plaquettes yields excellent agreement with experiment.
\end{abstract}

\maketitle
Josephson junction arrays (JJAs) exhibit rich dynamics, allowing the study of various quantum phase transitions \cite{R._Fazio_1994,d5ea6b1efacd4f92bfa307a684e7fcb9_2000,FAZIO2001235_2001}. 
Josephson networks have been extensively studied in the classical limit and are well described by the XY model \cite{PhysRevB.27.598_1983,PhysRevB.68.052502_2003,PhysRevB.102.094509_2020}. 
Applying a magnetic field frustrates the system, resulting in vortex lattice formation
\cite{PhysRevLett.51.1999_1983,PhysRevB.28.6578_1983}. 
Certain geometries, such as Lieb or dice lattices, are expected to exhibit flat energy bands when fully frustrated
\cite{PhysRevLett.81.5888, Korshunov2001vortex, Cataudella2003glassy, Korshunov2004Structure, Korshunov2005fluctuation, PhysRevLett.108.045306_2012, Leykam01012018_2018}, which is relevant for designing quantum coherent devices and synthetic quantum matter \cite{doi:10.1126/sciadv.adj7195,Neves_Wakefield_Fang_Nguyen_Ye_Checkelsky_2024,Chen_Li_Xie_Zhang_Lam_Tang_Lin_2024,PhysRevX.15.021091,Chen_Huang_Velkovsky_Ozawa_Price_Covey_Gadway_2025}.

Recent advances in semiconductor-superconductor hybrids \cite{PhysRevB.93.155402_2016,PhysRevMaterials.7.056201_2023} have enabled voltage-controlled
JJAs, allowing \textit{in-situ} control of quantum phase transitions, including superconductor-insulator and superconductor-metal transitions \cite{Bøttcher_Nichele_Kjaergaard_Suominen_Shabani_Palmstrøm_Marcus_2018,PhysRevB.110.L180502_2024,sasmal2025voltagetunedanomalousmetalmetaltransition}.
In the past few years, proximitized heterostructures based on two-dimensional electron gases (2DEGs) have been used to realize topological wires \cite{PhysRevLett.119.136803,PhysRevB.107.245423,Aghaee2025}, artificial Kitaev chains \cite{ten_Haaf_Wang_Bozkurt_Liu_Kulesh_Kim_Xiao_Thomas_Manfra_Dvir_et_al._2024}, Andreev molecules~\cite{Haxell_Coraiola_Hinderling_ten_Kate_Sabonis_Svetogorov_Belzig_Cheah_Krizek_Schott_et_al._2023,Matsuo_Imoto_Yokoyama_Sato_Lindemann_Gronin_Gardner_Nakosai_Tanaka_Manfra_et_al._2023}, supercurrent diodes \cite{Baumgartner_2021}, and Josephson elements with tunable current-phase relations~\cite{PhysRevLett.133.186303,PhysRevX.15.011021}.
However, effects of Andreev processes from the surrounding semiconductor remain mostly unexplored.

\begin{figure}[!b]
\includegraphics[width=\linewidth]{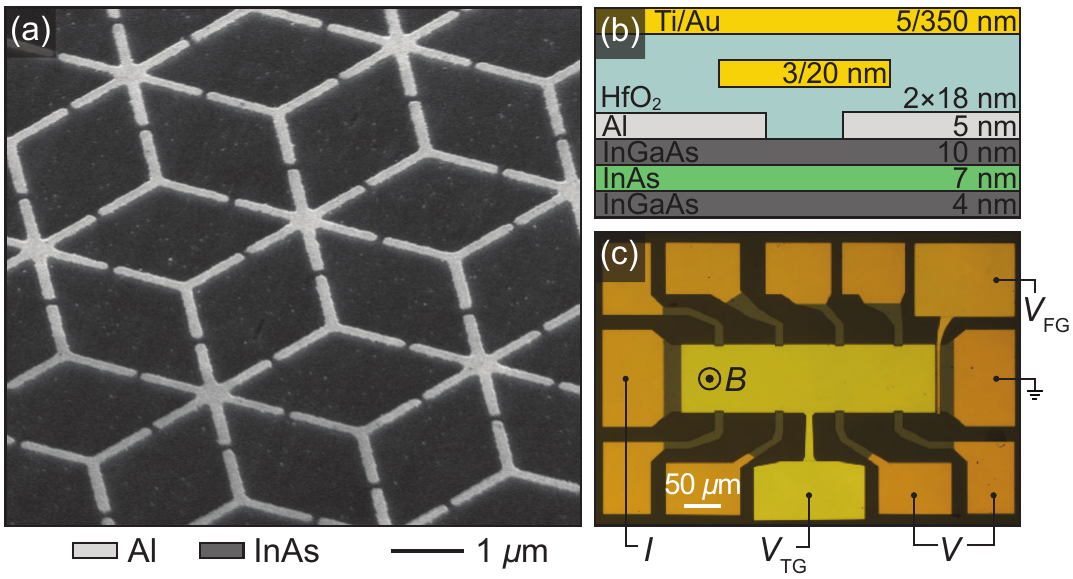}
\caption{(a)~Scanning electron micrograph of a hybrid Josephson junction array in dice-lattice
geometry showing superconducting Al islands (light) on a
semiconducting InAs heterostructure (dark), taken prior to gate fabrication.
(b)~Schematic cross section of a junction showing the dual gate configuration. The frame-gate (lower layer, kept at 0~V throughout the experiment) follows the Al geometry extending over the junctions, and tunes the inter-island coupling. The global top-gate (top layer) tunes the carriers in the plaquettes.
(c)~Optical image of device 1 overlaid with the measurement setup.
}
\label{fig:1}
\end{figure}

In this Letter, we investigate hybrid 2DEG-based Josephson arrays in the form of a dice lattice with two separate electrostatic gates controlling charge carrier density in the junctions and the plaquettes (Fig.~\ref{fig:1}). An applied perpendicular magnetic field, $B$, results in an intricate pattern of switching currents,~$I_{\rm SW}$, that is roughly periodic in frustration, $f$, or the flux, $\Phi$, threading each plaquette in units of $\Phi_0 = h/2e$, consistent with previous studies \cite{PhysRevLett.83.5102_1999, TESEI2006328, Beak2008superconducting}. We focus on the evolution of $I_{\rm SW}$ at integer $f$, examining how the periodicity in $f$ depends on the carrier population of the plaquettes. 

For depleted plaquettes, the array approximates a Josephson network of thin-wires connected by weak links, where $I_{\rm SW}(f)$ is periodic, with integer $f$ corresponding to integer winding of the superconducting phase around each loop without internal supercurrents~\cite{PhysRevLett.51.1999_1983}. In this thin-wire limit, the complex form of $I_{\rm SW}(f)$ depends on the particular lattice, with symmetry around $f=1/2$, repeating with period 1. 
For populated plaquettes, on the other hand, the system behaves like a proximitized two-dimensional film, where each integer $f$ imposes boundary conditions of $f$ phase twists on each plaquette. In particular, nonzero integer $f$ frustrates the proximity effect of the plaquettes---comparable to adding multiple vortices---and suppresses the proximity-induced gap. As a result, one would expect  $I_{\rm SW}(f)$ to decrease with increasing integer $f$. 
As presented below, this intuitive picture matches what is seen experimentally and in the model. 

The model developed here treats occupied plaquettes as long diffusive superconducting weak links with monotonic flux dependence. In this description, the plaquette contribution to $I_{\rm SW}(f)$ reflects the areal distribution of Andreev trajectories in the plaquettes. Although the junctions in the wire network are considerably smaller than the plaquettes, we  include their flux dependence in the model as well.

Devices were fabricated using a semiconductor-superconductor hybrid heterostructure 
\cite{PhysRevB.93.155402_2016,PhysRevMaterials.7.056201_2023}
consisting of a 7~nm InAs quantum well covered with a 5~nm epitaxial Al layer as described previously \cite{PhysRevLett.133.186303,PhysRevX.15.011021}. 
The Al was lithographically patterned into arrays of narrow superconducting islands forming a Josephson junction array in a dice lattice geometry, surrounded by plaquettes of uncovered semiconductor [Fig.~\ref{fig:1}(a)]. 
Two Ti/Au gates were then deposited, separated from the heterostructure and each other by a HfOx dielectric, grown by atomic layer deposition [Fig~\ref{fig:1}(b)]. The first layer, referred to as the frame-gate, follows the Al island geometry and extends over the junctions. A voltage applied to the frame-gate, $V_{\rm FG}$, tunes the Josephson couplings, $E_J$. The second layer is the global top-gate, that covers the entire array. The top-gate voltage, $V_{\rm TG}$, controls the charge carrier density in the plaquettes. The metallic frame-gate screens the junctions from the electric fields generated by the top-gate electrode.
The arrays were fabricated on a Hall bar geometry with leads and probes for sourcing and measuring electrical signals; see Fig.~\ref{fig:1}(c). 

Two devices, denoted 1 and 2, were measured and showed similar results. In the main text, we present data from device~1 with wider Al islands (250~nm), whereas data from device~2 with narrower islands (150~nm) are summarized in the Supplemental Material \cite{Supplement}.
Measurements were performed in a four-terminal configuration using standard lock-in techniques in a dilution refrigerator with a three-axis vector magnet, at a base temperature of 15~mK. For more details on device fabrication and measurement, see Supplemental Material \cite{Supplement}.

\begin{figure}[!t]
\includegraphics[width=\linewidth]{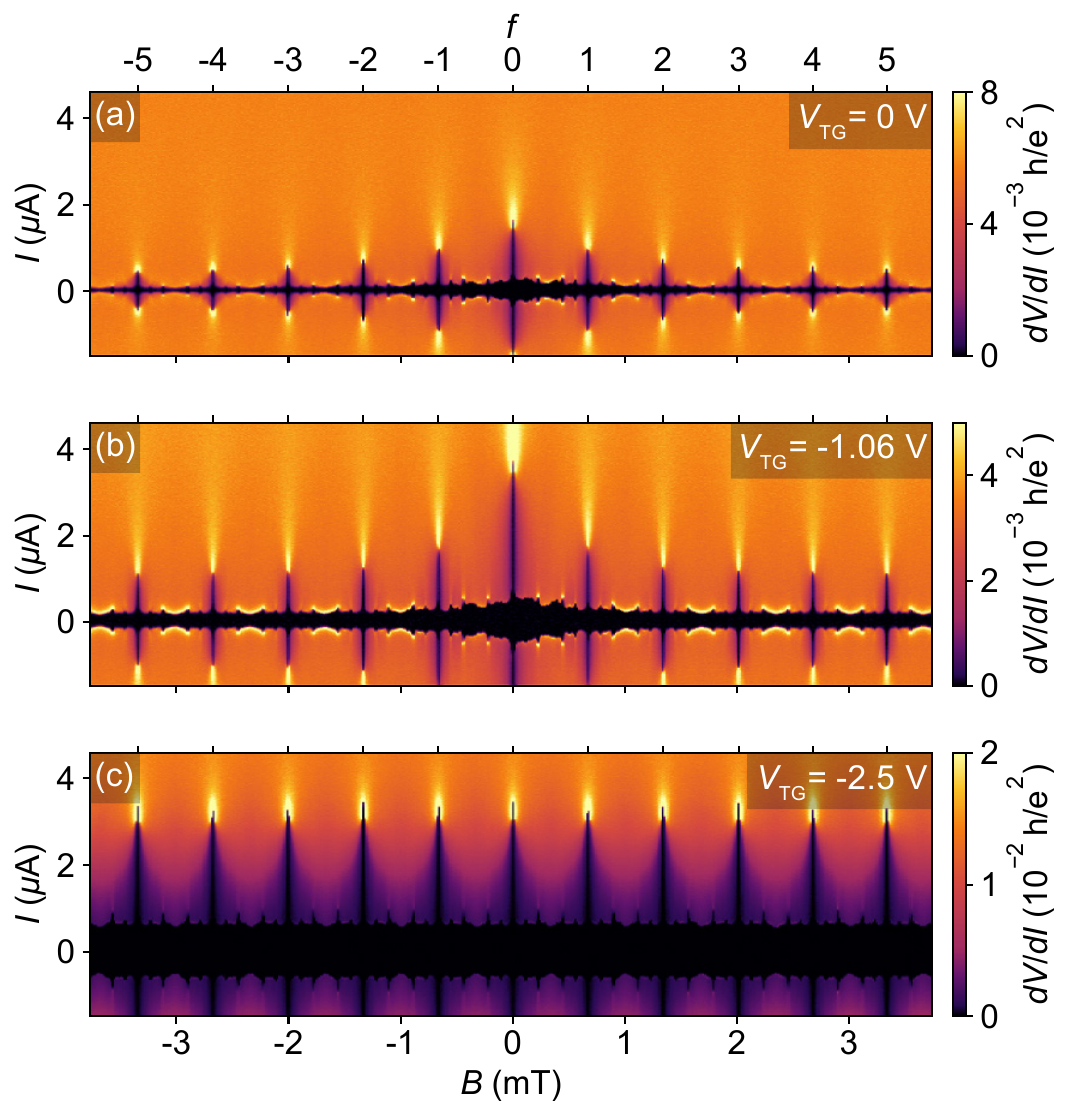}
\caption{ 
Differential resistance, $dV/dI$, as a function of current bias, $I$, and flux-threading magnetic field, $B$, taken at different top-gate voltages, (a)~$V_\text{TG}=0$, (b)~$-1.06$~V, and (c)~$-2.5$~V. The switching current, $I_\text{SW}$, is periodically modulated by $B$, showing sharp peaks at integer frustration values, $f=\Phi/\Phi_0$.
For populated plaquettes ($V_\text{TG}=0$ and $-1.06$~V), the envelope of $I_\text{SW}$ decays sharply with increasing $B$. For depleted plaquettes ($V_\text{TG}=-2.5$~V), the envelope is effectively flat throughout the measured range.
}
\label{fig:2}
\end{figure}

To characterize the switching current, $I_\text{SW}$, of the array, we measure its differential resistance, $dV/dI$,
as a function of source-drain current, $I$, and flux-threading magnetic field, $B$, at different top-gate voltages, $V_\text{TG}$; see Fig. \ref{fig:2}. We observe periodic sharp peaks in $I_\text{SW}$ at integer values of magnetic frustration $f$, typical of JJAs regardless of geometry \cite{PhysRevB.54.10081,PhysRevB.65.214504}.
Within each period, lower-amplitude peaks appear at commensurate fractions of flux, mostly visible at $f=\{1/6,\,1/3\,,2/3\,,5/6\}\pm n$, with integer $n$, consistent with the dice lattice geometry~\cite{PhysRevLett.83.5102_1999, TESEI2006328, Beak2008superconducting, Serret2002vortex}. A pronounced minimum is observed at $f=1/2$, consistent with Aharonov-Bohm caging in the fully frustrated dice-lattice \cite{PhysRevLett.81.5888}.
The main effect of tuning the $V_\text{TG}$ is a drastic modification of the envelope of the integer $I_\text{SW}$ peaks, while the sub-periodic frustration pattern remains qualitatively unchanged.
At $V_{\rm TG} = 0$, the envelope decays rapidly over a small magnetic field range, followed by a slower decay at larger $B$ [Fig.~\ref{fig:2}(a)]. At intermediate gate voltages, around $V_{\rm TG} = -1$~V, this two-stage behavior persists, but the central peak at $f = 0$ becomes considerably enhanced relative to the others [Fig.~\ref{fig:2}(b)]. At $V_{\rm TG} = -2.5$~V, the envelope flattens, and the switching current peaks become comparable at all integer frustrations [Fig.~\ref{fig:2}(c)]. 
A pronounced decay of the envelope with $B$ has been observed in previous experiments on JJAs with wide islands \cite{PhysRevB.108.134517_2023,reinhardt2025spontaneoussupercurrentsvortexdepinning_2025}. In contrast, the islands in our devices are narrower, and the observed change in the envelope can be associated with the array crossing from a proximitized two-dimensional film to a thin-wire Josephson network limit as a function of voltage-tuned carrier density in the plaquettes.

\begin{figure}[!t]
\includegraphics[width=\linewidth]{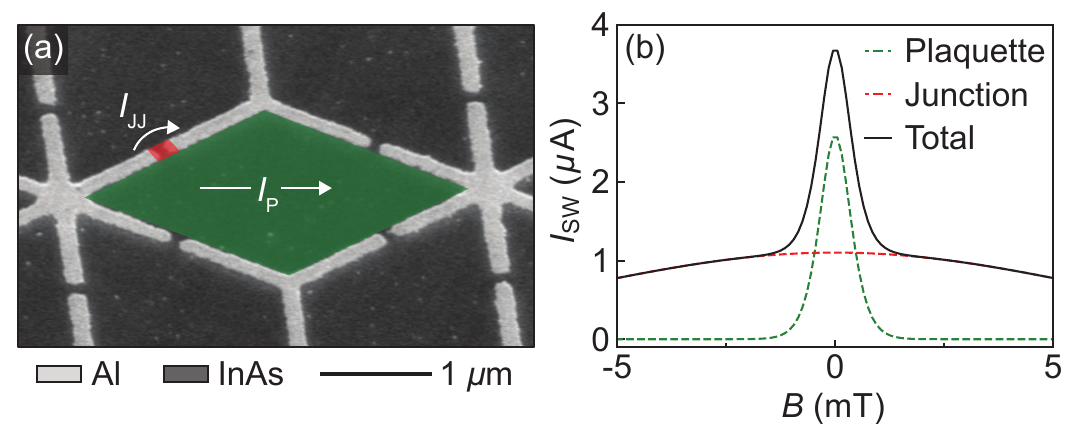}
\caption{(a) Enlarged electron micrograph of a single plaquette. (b) Switching current of the system, $I_{\rm SW}$, is modeled by considering contributions from the junctions (red) and plaquettes (green); see Eq. (\ref{model}). Because of the different areas, the contribution from the plaquettes is suppressed at higher fields, leaving the slow-decaying contribution from the junctions.
}
\label{fig:3}
\end{figure}

To describe the evolution of the envelope with $V_\text{TG}$, we develop a model with two independent supercurrent channels through the Josephson junctions and proximitized plaquettes; see Fig.~\ref{fig:3}(a).
We assume the junctions are narrow and short relative to the coherence length, and approximate the plaquettes as long diffusive Josephson junctions. With these considerations, the total switching current as a function of magnetic field can be expressed as
\begin{equation} \label{model}
    I_\text{SW}(B)=I_{\text{JJ}}\,e^{-\left(\frac{a_{\text{JJ}}B}{B_{\text{JJ}}}\right)^2}+I_\text{P}\,\frac{\frac{\pi}{\sqrt{3}}\frac{a_\text{P}B}{B_\text{P}}}{\sinh\frac{\pi}{\sqrt{3}}\frac{a_\text{P}B}{B_\text{P}}}\,,
\end{equation}
where $I_{\rm JJ}$ and $I_{\rm P}$ are the switching-current amplitudes of the junctions and plaquettes, $B_{\rm JJ} = \Phi_0/A_{\rm JJ}$ and $B_{\rm P} = \Phi_0/A_{\rm P}$ are the characteristic magnetic field scales associated with the junction and plaquette areas, $A_{\rm JJ}$ and $A_{\rm P}$, and $a_{\rm JJ}$ and $a_{\rm P}$ are the geometric scaling factors.
The first term in Eq.~(\ref{model}) reflects a Gaussian current profile typical for a narrow Josephson junction \cite{PhysRevB.3.3015_1971,PhysRevLett.99.217002_2007,PhysRevB.86.064510_2012,Li_2022}.
The second term follows from the theory of long diffusive SNS junctions with a sinusoidal current-phase relation \cite{montambaux2007interferencepatternlongdiffusive,PhysRevB.86.064510_2012,PhysRevB.87.024514_2013}.
Because $B_{\rm P} \ll B_{\rm JJ}$, the plaquette contribution decays more rapidly with field, while the junction contribution persists over a broader range; see Fig.~\ref{fig:3}(b). When the plaquettes are depleted, only the slowly-decaying junction contribution remains.

\begin{figure}[!t]
\includegraphics[width=\linewidth]{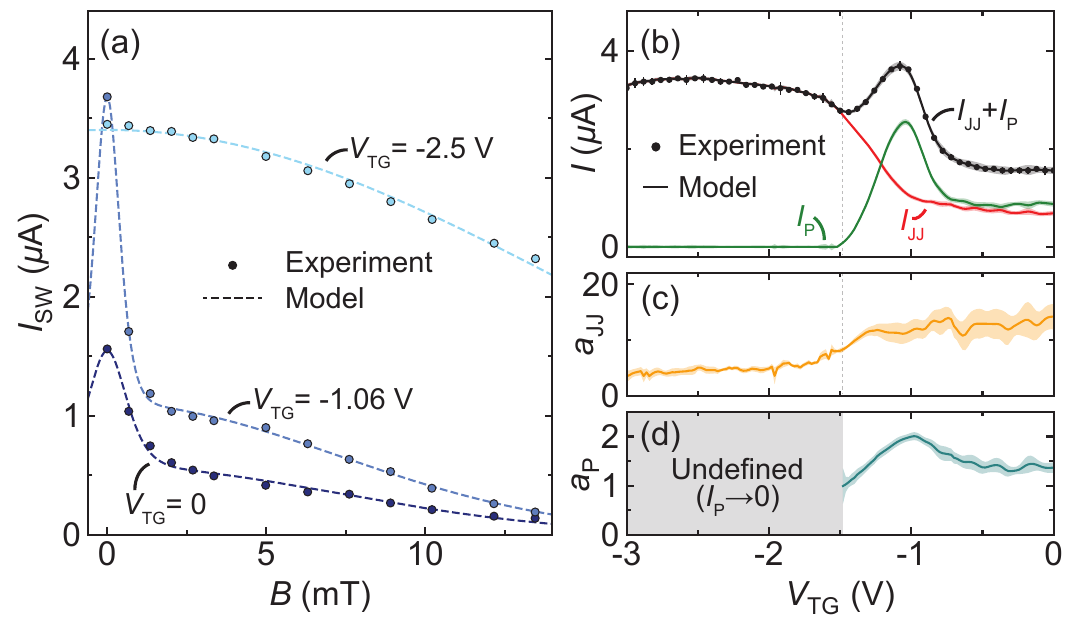}
\caption{ (a) Envelope of the switching current, $I_\text{SW}$, for three characteristic top-gate voltages, $V_\text{TG}$, as a function of flux threading magnetic field, $B$, measured at integer values of frustration, $f=\Phi/\Phi_0$. The dashed lines are fits to Eq. (\ref{model}).
(b) Extracted junction, $I_\text{JJ}$, and plaquette, $I_\text{P}$, switching-current amplitudes; for fitting details see Supplemental Material~\cite{Supplement}. The sum of the two contributions agree well with the experimental $I_\text{SW}$ at $B=0$. (c) and (d) Extracted dimensionless scaling factors for junction, $a_\text{JJ}$, and plaquette, $a_\text{P}$, areas as a function of $V_\text{TG}$. When plaquettes are depleted (around $V_\text{TG}=-1.5$ V), $a_\text{P}$ becomes undefined, since $I_\text{P}\to0$.
}
\label{fig:4}
\end{figure}

Fitting Eq.~(\ref{model}) to the extracted $I_{\rm SW}$ at integer $f$ gives good agreement in all three $V_{\rm TG}$ regimes shown in Fig.~\ref{fig:2}; see Fig.~\ref{fig:4}(a). The phenomenological model uses $I_{\rm JJ}$, $I_{\rm P}$, $a_{\rm JJ}$, and $a_{\rm P}$ as fit parameters that can be interpreted in terms of physically relevant quantities, whereas $B_{\rm JJ}$ and $B_{\rm P}$ are defined by the lithographic areas of the corresponding junctions.
When the plaquettes are populated ($V_{\rm TG} = 0$ and $-1.06$~V), the envelope exhibits a clear two-component decay—a sharp drop at low field followed by a slower decay at higher field. At intermediate gate voltage ($V_{\rm TG} = -1.06$~V), the $f~=~0$ peak becomes visibly sharper. When the plaquettes are depleted ($V_{\rm TG} = -2.5$~V), the envelope transitions to a single-component decay. We note that the amplitude of the slow-decaying component increases with depletion, indicating that dissipation plays an important role in determining the $I_{\rm SW}$ magnitude of the array.

To better understand the evolution of the parameters, we perform the fit across a range of $V_{\rm TG}$ values; see Fig.~\ref{fig:4}(b–d). The plaquette amplitude $I_{\rm P}$ exhibits a pronounced peak near $V_{\rm TG} = -1.06$~V and is suppressed as the plaquettes are depleted. 
In contrast, $I_{\rm JJ}$ remains roughly constant throughout the populated regime, then increases sharply as $I_{\rm P} \rightarrow 0$. The sum of the two amplitudes follows the measured switching current at $f=0$. 
We note that the higher-frustration peaks predominantly follow the junction contribution (see Supplemental Material \cite{Supplement}). The junction scaling factor exhibits a step-like behavior, decreasing from $a_{\rm JJ} \gtrsim 10$, when the plaquettes are populated, to around 5 in the depleted regime. 
The plaquettes scaling factor displays a peak with $a_{\rm P} \gtrsim 2$ that aligns with the maximum in $I_{\rm P}$ and approaches unity as the plaquettes are depleted.

We interpret the behavior of $I_{\rm P}$ as reflecting the gate-voltage-dependent mobility of the 2DEG in the plaquettes. As $V_{\rm TG}$ is decreased, the carrier density reduces monotonically, which in turn suppresses inter-subband and surface scattering prior to depletion. This gives rise to a mobility peak, characteristic of shallow 2DEG heterostructures~\cite{PhysRevMaterials.2.104602_2018, PhysRevMaterials.7.056201_2023,sasmal2025voltagetunedanomalousmetalmetaltransition} (see Supplemental Material \cite{Supplement}).
Higher mobility corresponds to a longer mean free path and results in a larger $I_{\rm P}$. 

The observed increase in $I_{\rm JJ}$ with depletion suggests that dissipation plays an important role in shaping the proximity effect. This behavior is consistent with the inverse proximity effect, where coupling to a normal reservoir suppresses superconducting correlation~\cite{RevModPhys.36.225_1964, PhysRevB.87.180504_2013, PhysRevLett.121.037703_2018, PhysRevX.8.031040_2018, de_Moor_2018,10.1063/5.0215522_2024}.
In this context, populated plaquettes lead to a lower probability of coherent Andreev processes, which weakens the proximity effect \cite{J_A_Melsen_1997,PhysRevB.94.020501}. As carriers are depleted, quasiparticles become confined beneath the Al islands, dissipation is removed, and $I_{\rm JJ}$ increases.

We interpret $a_{\rm JJ}$ in terms of the Josephson penetration depth $\lambda_{\rm J}$, which sets the effective junction area via $A_{\rm JJ}^{\rm eff}~=~w_{\rm JJ} \lambda_{\rm J}$, with $\lambda_{\rm J}~\approx~a_{\rm JJ}
\ell_{\rm JJ}$, where $w_{\rm JJ}$ is the lithographic width of the Josephson junction and $\ell_{\rm JJ}$ is its length.
Using $\ell_{\rm JJ}~=~(120 \pm 10)$~nm we find $\lambda_{\rm J} = (1.7 \pm 0.3)$~$\mu$m in the populated regime ($V_{\rm TG} = 0$) and $\lambda_{\rm J}~=~(0.6 \pm 0.1)~\mu$m in the depleted
regime ($V_{\rm TG} = -2.5$~V).
This can be compared to the expectation for kinetic-inductance-dominated coplanar junctions, with the Josephson
penetration length given by $\lambda_{\rm J}~=~\sqrt{ \Phi_0 t_{\rm JJ} w_{\rm JJ}^2 / 4\pi \mu_0 \lambda_L^2 I_{\rm JJ}}$~\cite{10.1063/1.117568}, where $t_{\rm JJ} = (17 \pm 5)$~nm is the junction thickness, taken as the combined thickness of the top
barrier and quantum well layers [see Fig. \ref{fig:1}(b)], $w_{\rm JJ} = (250 \pm 5)$~nm, and $\lambda_L = (200 \pm 60)$~nm is the dirty-limit London penetration depth accounting for the thinness of Al \cite{PhysRevB.101.060507,10.1088/1361-6668/adf360}.
This estimate yields $(2.3 \pm 0.8)~\mu$m and $\lambda_{\rm J} = (1.0
\pm 0.3)$~$\mu$m in the populated and depleted regimes, respectively, which agrees within uncertainties with values deduced from the fitted $a_{\rm JJ}$.
We note that the effective junction width might be smaller than $w_{\rm JJ}$ due to the fringing electric field from the top-gate, which would explain the slightly higher $\lambda_{\rm J}$ estimated using the coplanar-junction equation.

Finally, the evolution of $a_{\rm P}$ can be understood as reflecting the change in the nature of the transport regime in the plaquettes.
The mean free path, $l_{e}$, computed from
the gate-dependent mobility, $\mu$, and carrier density, $n_{e}$, (see Supplemental Material \cite{Supplement}), increases from roughly 230~nm at $V_{\rm TG} = 0$ to around 640~nm near the peak in $a_{\rm
P}$ at $V_{\rm TG} = -1.06$~V. The corresponding induced coherence length in the dirty limit can be estimated as $\xi_{\rm P} \approx 0.855\sqrt{\hbar l_{e}v_{\rm F} / \pi \Delta}~\approx~\sqrt{3\,\hbar^3\mu n_{e}/2 \Delta m^* \vert e\vert}$, where $v_{\rm F}$ is the density-dependent Fermi velocity, $\Delta = 200\,\mu\text{eV}$ is
the induced superconducting gap, and $m^*=0.023\,m_{e}$ is the effective mass, with electron mass, $m_{e}$~\cite{Tinkham_2004}. This yields a moderate increase of $\xi_P$ from $\sim 600$ to 700~nm over the same gate-voltage range. Compared to the lithographic plaquette
length ($\sim 3.5~\mu \rm m$), this indicates that transport in this regime deviates from the long junction limit (see Supplemental Material \cite{Supplement}). 

In summary, we have examined the magnetic-field envelope of switching current in InAs/Al dice-lattice Josephson junction arrays.
We find that the envelope depends strongly on the voltage-controlled carrier density in the plaquettes of proximitized semiconductor surrounding the junctions, indicating a crossover from a proximitized film to a network of weak links. This behavior is well captured by a two-channel model incorporating parallel contributions from junction and plaquette supercurrents. Fitting the model to the data yields the gate-voltage dependence of the transport parameters, providing quantitative insight into the proximity effect in frustrated Josephson junction arrays.\\

\textit{Acknowledgments -} We thank A. C. C. Drachmann for assistance with device
fabrication and C. Kvande, M. R. Lykkegaard, J. Zhao, and L. Casparis for discussions. We acknowledge support from research grants (Projects No. 43951 and
No. 53097) from VILLUM FONDEN, the Danish National
Research Foundation, and the European Research Council
(Grant Agreement No. 856526).

\bibliography{bibliography.bib}

\end{document}


\title{Supplemental Material:\\From two dimensions to wire networks in a dice-lattice Josephson array}
\author{J. D. Bondar}
\affiliation{Center for Quantum Devices, Niels Bohr Institute,
University of Copenhagen, 2100 Copenhagen, Denmark}
\affiliation{Department of Physics and Astronomy, Purdue University, West Lafayette, Indiana 47907, USA}
\author{L. Banszerus}
\affiliation{Center for Quantum Devices, Niels Bohr Institute, University of Copenhagen, 2100 Copenhagen, Denmark}
\affiliation{Faculty of Physics, University of Vienna, 1090 Vienna, Austria}
\author{W. Marshall}
\affiliation{Center for Quantum Devices, Niels Bohr Institute,
University of Copenhagen, 2100 Copenhagen, Denmark}
\affiliation{Department of Physics, University of Washington, Seattle, Washington 98195, USA}
\author{T. Lindemann}
\affiliation{Department of Physics and Astronomy, Purdue University, West Lafayette, Indiana 47907, USA}
\author{T.~Zhang}
\affiliation{Department of Physics and Astronomy, Purdue University, West Lafayette, Indiana 47907, USA}

\author{M. J. Manfra}
\affiliation{Department of Physics and Astronomy, Purdue University, West Lafayette, Indiana 47907, USA} 
\affiliation{
Purdue Quantum Science and Engineering Institute, Purdue University,  West Lafayette, Indiana 47907, USA}
\affiliation{Elmore Family School of Electrical and Computer Engineering, Purdue University, West Lafayette, Indiana 47907, USA}
\affiliation{School of Materials Engineering, Purdue University, West Lafayette, Indiana 47907, USA}
\author{C. M. Marcus}
\affiliation{Center for Quantum Devices, Niels Bohr Institute,
University of Copenhagen, 2100 Copenhagen, Denmark}
\affiliation{Materials Science and Engineering, and Department of Physics, University of Washington, Seattle, Washington 98195, USA
}
\author{S. Vaitiekėnas}
\affiliation{Center for Quantum Devices, Niels Bohr Institute,
University of Copenhagen, 2100 Copenhagen, Denmark}
\date{\today}

\maketitle
\section*{Sample Preparation}
The devices were fabricated on a semiconductor-superconductor hybrid heterostructure. The same heterostructure has been used in previous works \cite{PhysRevLett.133.186303,PhysRevX.15.011021}. The III-V semiconductor stack was grown on a semi-insulating InP substrate by molecular beam epitaxy (MBE). It consists of
a 7~nm thick InAs quantum well, encapsulated between 10~nm (4~nm) thick In$_{0.75}$Ga$_{0.25}$As top (bottom) barrier, followed
by two monolayers of GaAs capping, which acts as an etch stop. The 5~nm Al film
was grown \it{in situ} \rm without exposing the heterostructure
to ambient conditions.

\begin{figure}[!b]
\includegraphics[width=\linewidth]{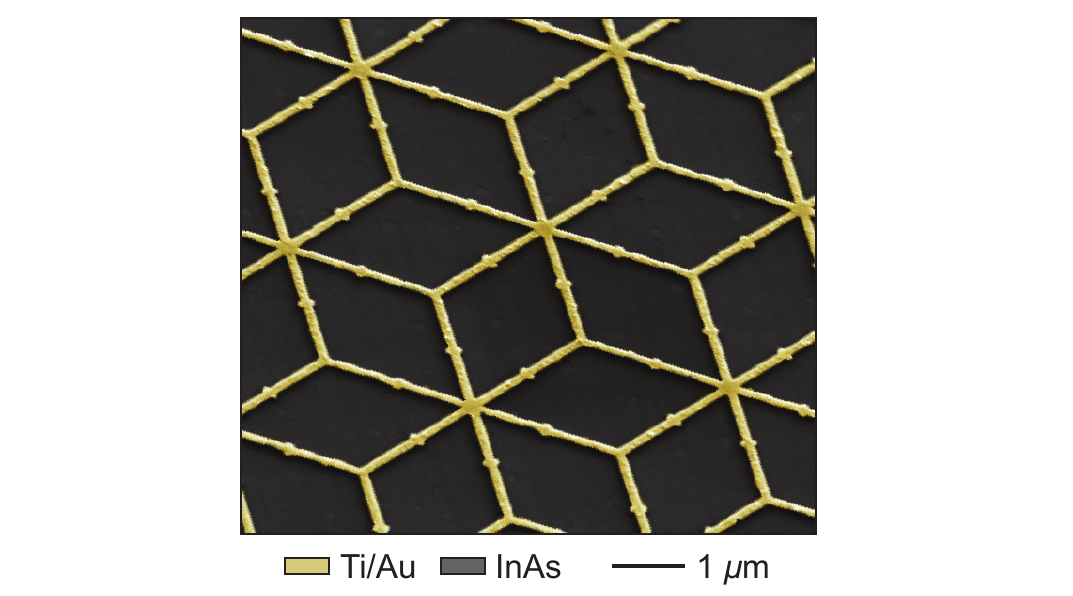} 
\caption{\justifying Scanning electron micrograph of the Ti/Au frame-gate (yellow) patterned to follow the Josephson junction array in dice-lattice geometry. 
The gate overlaps the junctions for tuning inter-island coupling and screening the junctions from the electric field of the top-gate (not shown).
}
\label{fig:S1}
\end{figure}

The devices were patterned using standard electron beam lithography. The Al was selectively  etched using Transene Aluminum etch type~D
at $50\,^\circ$C for 5 seconds. 
The heterostructure
was structured by a chemical wet etch using (220:55:3:3
H$_2$O:C$_6$H$_8$O$_7$:H$_3$PO$_4$:H$_2$O$_2$) into Hall bar geometries with a length of 570~$\mu m$ and width of 150~$\mu$m. The array of $165\times50$ dice-lattice unit cells (3 plaquettes per unit cell) was patterned on top of the bars.
A layer of HfOx (18~nm) dielectric was grown using atomic layer deposition before each metalization of gate electrodes.
Both, the frame-gate, Ti/Au (3/20~nm), and then the global top-gate, Ti/Au (5/350~nm) [see Fig.~\ref{fig:S1}], were deposited using electron beam evaporation.
The frame gates of both devices were kept at 0~V throughout the experiment.

\begin{figure}[!b]
\includegraphics[width=\linewidth]{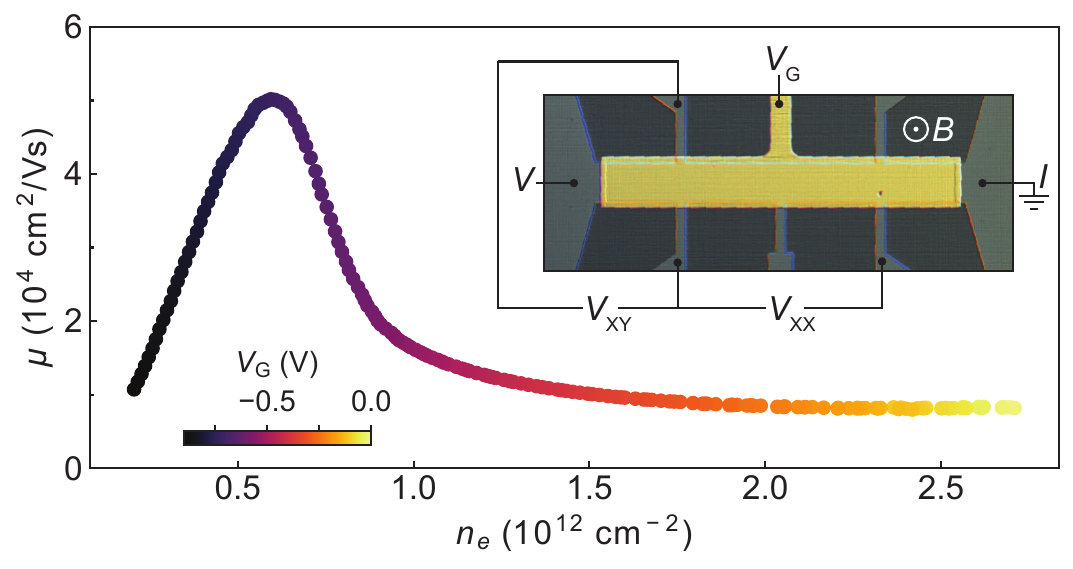} 
\caption{\justifying Charge carrier mobility, $\mu$, as a function of carrier density,
$n_{e}$, measured for a Hall bar with the Al film removed. The color shows
the applied top-gate voltage, $V_\text{G}$. Inset: Optical image of
the Hall-bar device, overlaid with the measurement setup.
}
\label{fig:S2}
\end{figure}

Characterization of the electrical properties of the two-dimensional electron gas was done using a top-gated Hall bar with the Al film removed (see Fig.~\ref{fig:S2} inset). A peak mobility of $\mu=50000$~$\rm{cm}^2/$Vs was measured at a charge carrier density $n_{e}=0.6\times10^{12}$~$\rm cm^{-2}$; see Fig.~\ref{fig:S2}. 
For $V_{\rm G}=0$, a mobility of $\mu=8500$~$\rm{cm}^2/$Vs and \mbox{$n_{e}=2.7\times10^{12}~{\rm cm^{-2}}$} were measured. 
These values of $\mu$ and $n_{e}$ are used in the main text to compute the electron mean free path, $l_{e}=\mu\hbar\sqrt{2\pi n_{e}}/|e|$.

\section*{Measurement Details}
Transport measurements were performed in a dilution refrigerator at a base temperature of 15 mK. 
Two devices were measured in parallel. Standard low-frequency ac lock-in measurement techniques were used at $f= 27.7$~Hz for device 1 and $f = 31.1$~Hz for device 2. 
All lines were filtered at cryogenic temperatures using a two-stage RC and LC filter with a cut-off frequency of 65 kHz.
Gate lines were
filtered using~16 Hz low-pass filters at room temperature. 
The bias was applied to the device via a home-built voltage divider (1:1000) and a load resistor $R_{\rm L} = 100$~k$\Omega$. An ac current excitation of 10~nA was used.
The currents were amplified using a current-to-voltage converter with a gain of $10^6$ for both devices. Voltages were amplified using preamplifiers with a gain of $10^3$ for both devices.

\section*{Additional Data}
\begin{figure}[!b]
\includegraphics[width=\linewidth]{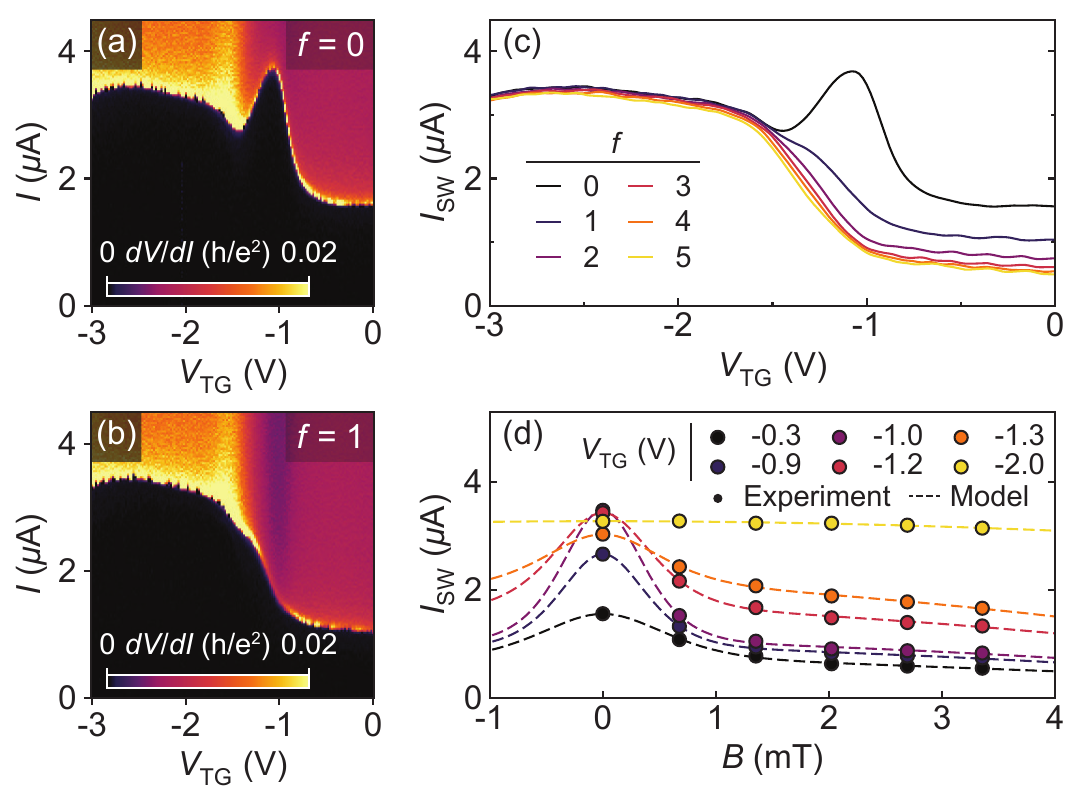} 
\caption{\justifying (a) Differential resistance, $dV/dI$, as a function of current bias, $I$, and top-gate voltage, $V_{\rm TG}$, measured for device 1 at magnetic frustration, $f=0$. (b) Same as (a) but at $f=1$. (c)~Extracted switching current, $I_{\rm SW}$, as a function of $V_{\rm TG}$ for integer $f$ from 0 to 5. (d)~Envelopes of $I_{\rm SW}$ for selected $V_{\rm TG}$, as a function of flux threading magnetic field, $B$, measured at integer values of $f$. The dashed lines are fits using the main-text Eq.~(1).}
\label{fig:S3}
\end{figure}
\subsection*{Peak at $f=0$}

Differential resistance, $dV/dI$, as a function of current bias, $I$, measured for $f=0$ and $f=1$, show a qualitatively different evolution with the top-gate voltage, $V_{\rm TG}$; see Fig.~\ref{fig:S3}(a) and \ref{fig:S3}(b). 
Specifically, there is a clear peak for $f=0$ around $V_{\rm TG}=-1$~V, that is missing for $f=1$. In fact, all higher integer frustration values do not show the peak and predominantly follow the gate evolution of $I_{\rm JJ}$; see Fig.~\ref{fig:S3}(c) and the main-text Fig.~4(b). 
This can be understood by considering the difference between the characteristic magnetic-field scales of the junctions, $B_{\rm JJ}$, and the plaquettes, $B_{\rm P}$. Since the peak in $I_{\rm SW}$ can be related to the mobility of the plaquettes and $B_{\rm P}~\ll~B_{\rm JJ}$, the plaquette contribution is rapidly suppressed with $B$, leaving only the junction contribution. As a result the $I_{\rm SW}$ does not appear at higher integer frustration values, with only minor contributions to the $f=1$ and $f=2$ traces; see Fig.~\ref{fig:S3}(c).

\subsection*{Fitting Details}

To obtain the gate dependence of the fit parameters in the model [see Eq. (1) in the main text], we extract the $I_{\rm SW}$ as a function of $V_{\rm TG}$ using thresholding for integer frustration peaks ranging from $f=0$ to $f=5$; see Fig.~\ref{fig:S3}(c). We note that an offset in $B$ of 39.89~$\mu$T was corrected before taking the data. We further note that for the fitting to converge, a small numerical offset of $10^{-20}$~T (15 orders of magnitude lower than the characteristic field scale $B_{\rm P}$) is added to the $f=0$ data point.
\begin{figure}[!b]
\includegraphics[width=\linewidth]{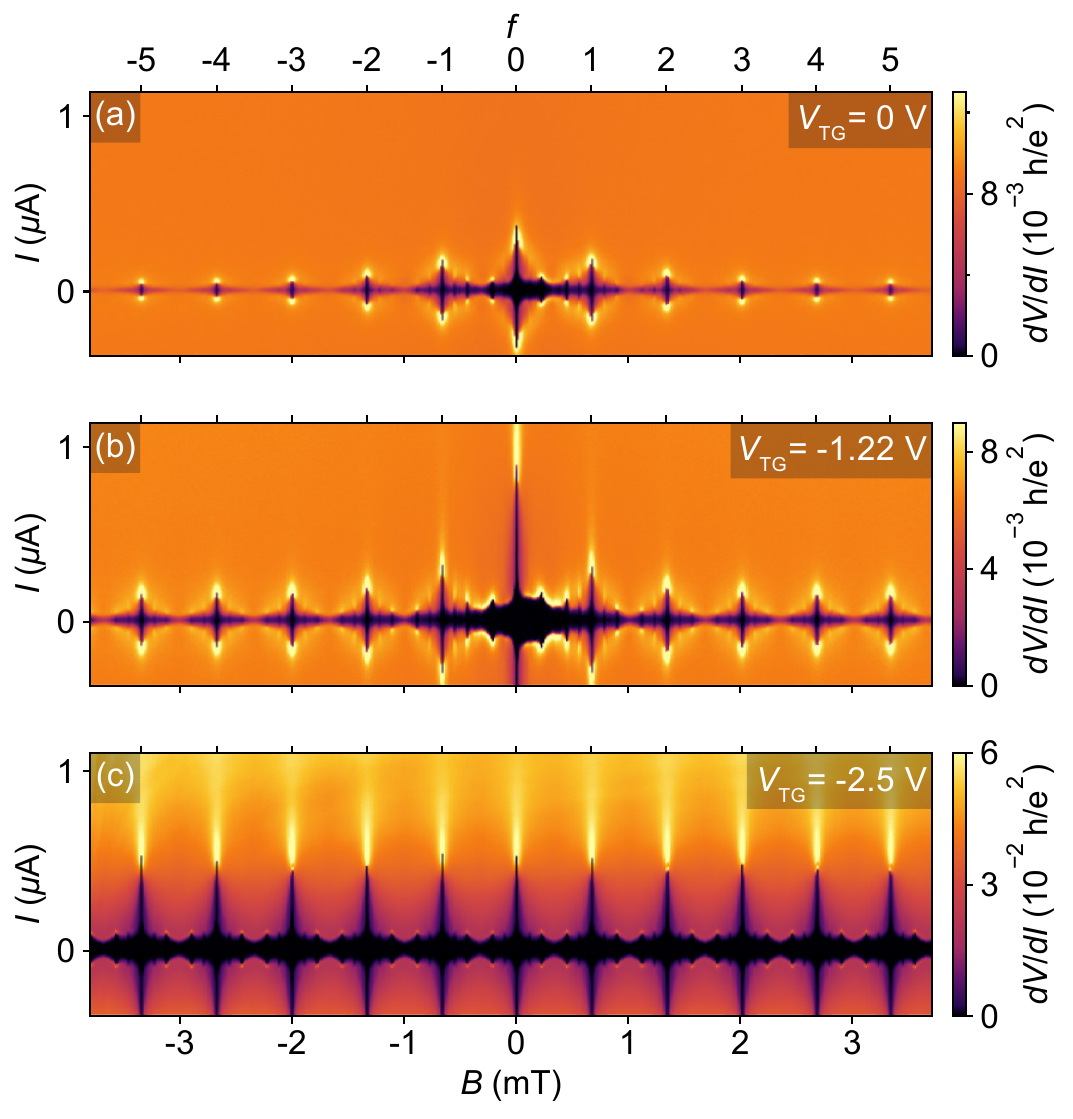} 
\caption{\justifying Differential resistance, $dV/dI$, measured for device 2 as a function of current bias, $I$, and flux-threading magnetic field, $B$, taken at different top-gate voltages, (a)~$V_\text{TG}=0$, (b)~$-1.22$~V, and (c)~$-2.5$~V.}
\label{fig:S4}
\end{figure}
Before fitting the envelopes, we apply a Gaussian averaging ($\sigma=1.5$) to the extracted $I_{\rm SW}(V_{\rm TG})$ at all six $f$ values. Some example envelopes and the corresponding fits to the main-text Eq. (1) are shown in Fig.~\ref{fig:S3}(d).
We constrain $I_{\rm P}$ close to 0 when fitting past plaquette depletion to allow for the accurate fitting of $I_{\rm JJ}$; without constraints we still observe $I_{\rm P}\to 0$ past depletion, though with larger uncertainties. We also constrain $a_{\rm JJ}$ differently in the populated and depleted plaquette regimes based on what results in accurate fits. We observe good fits of the model to the extracted $I_{\rm SW}$.
We note that for this analysis we used fewer $f$ values, compared to the main-text Fig.~4(a), which still yields good estimates for $I_\text{JJ}$, $I_\text{P}$, and $a_\text{P}$; however, $a_{\rm JJ}$ is generally overestimated in the populated plaquettes regime ($V_\text{TG}>-1.5$ V) due to only a few points in the $B$-range where the $I_{\rm SW}$ envelope is dominated by the junction contribution. In the depleted plaquettes regime ($V_\text{TG}<-1.5$ V) the analysis gives more accurate $a_\text{JJ}$ values.

\subsection*{Second Device}
A second device with narrow islands $(150\pm10)$~nm was measured parallel to the main device. We note that a $B$-offset of 73.79~$\mu$T was corrected for data from device 2. Differential resistance, $dV/dI$, measured for device 2, as a function of source-drain current, $I$, and flux-threading magnetic field, $B$, at different top-gate voltages, $V_{\rm TG}$, shows a qualitatively similar behavior to device 1; see Fig.~\ref{fig:S4}. Specifically, device 2 shows the same hallmark features of the dice-lattice array, and the drastic modification of the $I_{\rm SW}$ envelope with varying $V_{\rm TG}$. Quantitatively, device 2 shows a reduced $I_{\rm SW}$ amplitude, but a relatively sharper $I_{\rm SW}$ peak at $f=0$ ($V_{\rm TG}=-1.22$~V) compared to device 1.

\begin{figure}[!t]
\includegraphics[width=\linewidth]{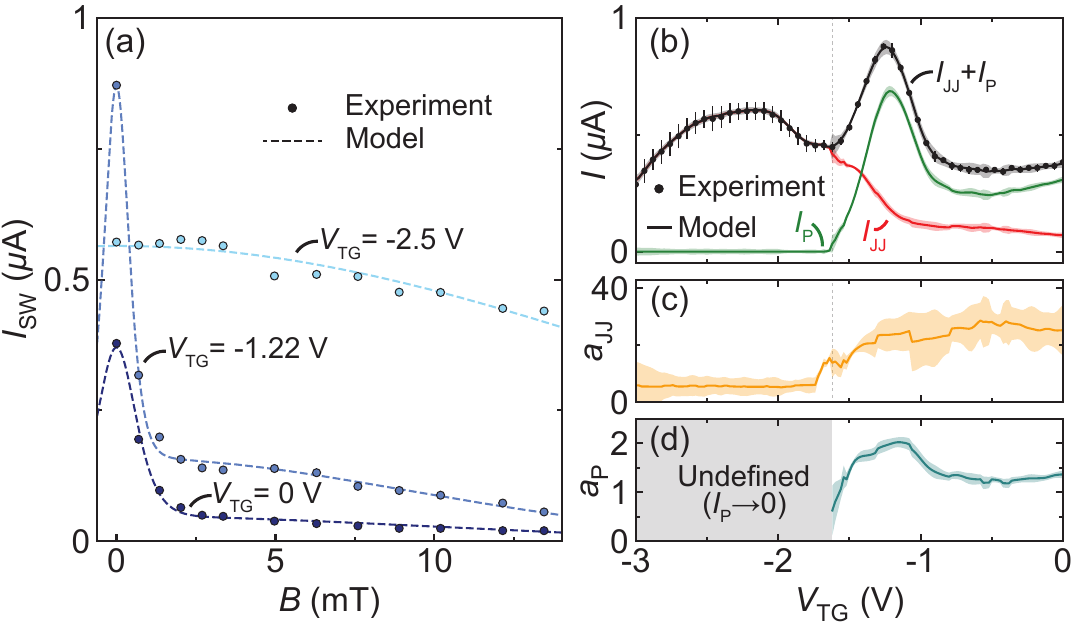} 
\caption{\justifying (a) Envelope of the switching current, $I_\text{SW}$, taken for device 2 for three characteristic top-gate voltages, $V_\text{TG}$, as a function of flux threading magnetic field, $B$, measured at integer values of frustration. The dashed lines are fits to the main-text Eq. (1). (b) Junction, $I_\text{JJ}$, and plaquette, $I_\text{P}$ switching-current amplitudes as a function of $V_\text{TG}$. (c) and (d) Extracted dimensionless scaling factors for junction, $a_\text{JJ}$, and plaquette, $a_\text{P}$, areas as a function of $V_\text{TG}$.}
\label{fig:S5}
\end{figure}

Fitting the main-text Eq. (1) to the $I_{\rm SW}$ envelope in all three distinct $V_{\rm TG}$ regimes shown in Fig.~\ref{fig:S4} describes the data well; see Fig.~\ref{fig:S5}(a). The voltage dependence of the fit parameters, $I_{\rm JJ}$, $I_{\rm P}$, $a_{\rm JJ}$, and $a_{\rm P}$, deduced from the fits across a range of $V_{\rm TG}$ values, reproduce the behavior observed in device 1; see Figs. \ref{fig:S5}(b) to \ref{fig:S5}(d). We attribute the decrease in $I_{\rm JJ}$ around $V_{\rm TG}=-3$~V to Josephson junctions being narrower and hence electrostatically effected by the top-gate.

\section*{Long Vs. Short Regimes}
\begin{figure}[!t]
\includegraphics[width=\linewidth]{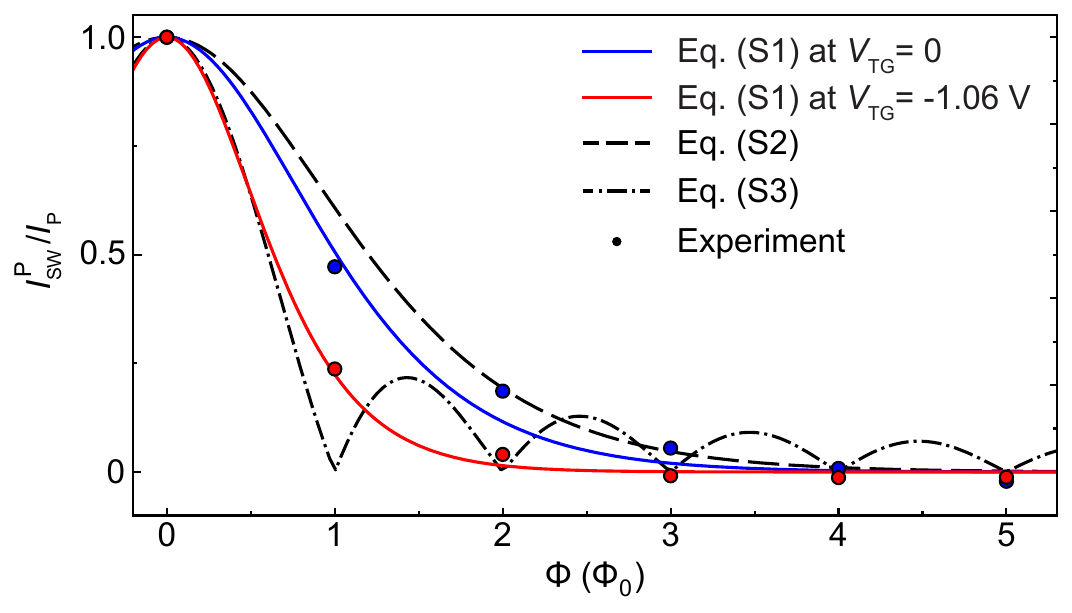} 
\caption{\justifying Normalized switching current, $I_{\rm SW}$, extracted from integer frustration peaks ranging from $f=0$ to 5, as a function of flux, $\Phi$, through the plaquettes, that are considered as Josephson junctions. The experimental points are normalized using Eq.~(\ref{norm}). The solid lines are normalized $I_{\rm SW}^{\rm P}$ curves deduced by fitting the main-text Eq. (1) to a wider range of integer $f$ values at $V_{\rm TG}=0$ (blue) and at the plaquette mobility peak at $V_{\rm TG}=-1.06$~V (red). The black curves are the theoretical switching currents  for long diffusive (long dashes) and short ballistic (dot dashes) junctions.}
\label{fig:S6}
\end{figure}

The extracted induced superconducting coherence length, $\xi_{\rm P}$, increases when the plaquette mobility peaks. This is reflected in the sharpening and narrowing of the plaquette contribution to the total $I_{\rm SW}$ (see fitted blue and red envelopes in Fig.~\ref{fig:S6}), which can be isolated by Eq.~(\ref{norm})
\begin{equation} \label{norm}
    I_\text{SW}^{\rm P}/I_{\rm P}=\frac{I_{\rm SW}-I_{\rm SW}^{\rm JJ}}{\max\left(I_{\rm SW}-I_{\rm SW}^{\rm JJ}\right)}\,,
\end{equation}
where $I_{\rm SW}^{\rm JJ}$ is the junction contribution to the total $I_{\rm SW}$.  

To better understand this behavior we look at the two extremes of the possible transport regimes for the plaquettes. One of the extremes is a long and diffusive (LD) junction, whose (normalized) switching current is given by \cite{montambaux2007interferencepatternlongdiffusive} 
\begin{equation} \label{diffusive}
    I_\text{SW}^{\rm LD}=\frac{\frac{\pi}{\sqrt{3}}\frac{\Phi}{\Phi_0}}{\sinh\frac{\pi}{\sqrt{3}}\frac{\Phi}{\Phi_0}}\,,
\end{equation}
where $\Phi$ is the flux through the junction and $\Phi_0=h/2e$ is the flux quanta. Note that Eq. (\ref{diffusive}) is similar to the plaquette contribution, $I_{\rm SW}^{\rm P}$, used in main-text Eq.~(1) but does not include $a_{\rm P}$.
The other extreme is a short and ballistic (SB) junction with the (normalized) switching current given by \cite{PhysRev.140.A1628}
\begin{equation} \label{ballistic}
    I_\text{SW}^{\rm SB}=\left|\frac{\sin\pi\Phi/\Phi_0}{\pi\Phi/\Phi_0}\right|\,.
\end{equation}
Plotting Eqs. (\ref{diffusive}) and (\ref{ballistic}) together with the experimental data points [normalized using Eq. (\ref{norm})] reveals that the points lie in-between $I_{\rm SW}^{\rm LD}$ and $I_{\rm SW}^{\rm SB}$ curves; see Fig.~\ref{fig:S5}. For populated plaquettes ($V_{\rm TG}=0$), the $I_{\rm SW}$ envelope is closer to $I_{\rm SW}^{\rm LD}$ [Eq. (\ref{diffusive})], whereas at the mobility peak ($V_{\rm TG}=-1.06$~V), the data points are slightly closer to $I_{\rm SW}^{\rm SB}$ [Eq. (\ref{ballistic})]. This indicates that as plaquettes approach the mobility peak, they edge from a long toward a short junction limit which is quantified by the increase in $a_{\rm P}$. 
In our model, $a_{\rm P} > 1$ accounts for the deviation from the long-junction limit, where the coherence length is not much smaller than ($\ll$), but comparable to ($\lesssim$) the dimensions of the plaquettes.
We note that just before the plaquette depletion, we observe $a_{\rm P}\approx1$, indicating that the plaquettes approach the idealized long junction limit. 

\bibliography{bibliography.bib}